\documentclass[11pt]{article}

\usepackage[T1]{fontenc}
\usepackage{float}
\usepackage{textcomp}
\usepackage{mathpazo}
\usepackage{verbatim}		
\usepackage{amsmath,amssymb}		
\usepackage{latexsym}		
\usepackage{marvosym}		
\usepackage[]{hyperref}		
\usepackage[T1]{fontenc}	
\usepackage[]{graphicx}	
\usepackage{endnotes}
\usepackage{multirow}
\usepackage[round]{natbib}
\usepackage[usenames,dvipsnames]{xcolor}
\usepackage{tikz}		%
\usetikzlibrary{plotmarks}
\usetikzlibrary{arrows}
\usepackage{color}
\usepackage{amssymb}
\usepackage{amsmath}
\usepackage{placeins}
\usepackage{caption} 
\usepackage{wrapfig}
\usepackage{bm}
\usepackage{comment}
\usepackage{subcaption} 
\usepackage{booktabs}

\makeatletter

\usepackage[top=2.54cm, bottom=2.54cm, left=2.54cm, right=2.54cm]{geometry}

\renewcommand{\tablename}{\footnotesize \textsc{Table}}
\renewcommand{\fnum@table}{\footnotesize \scshape \tablename~\thetable}

\renewcommand{\figurename}{\footnotesize \textsc{Figure} }
\renewcommand{\fnum@figure}{\footnotesize \scshape \figurename~\thefigure}

\newcommand{\newsupport}[2]{      
    \newenvironment{#1}
    {\begin{trivlist}
        \setlength{\parskip}{0mm}
        \footnotesize
        \item[\hskip \labelsep {\bfseries #2}]}
        {\QED
        \normalfont
        \setlength{\parskip}{\baselineskip}
        \end{trivlist}}}
	
\newcommand{\QED}{								
			\nobreak \ifvmode \relax \else
      \ifdim\lastskip<1.5em \hskip-\lastskip
      \hskip1.5em plus0em minus0.5em \fi \nobreak
      \vrule height0.5em width0.5em depth0.25em\fi}

    \newsupport{support}{Support}	    

    {\setcounter{table}{0}\renewcommand{\thetable}{\Alph{table}}
    \setcounter{figure}{0}\renewcommand{\thefigure}{\Alph{figure}}
 		\setcounter{subsection}{0}
 		}{}

		\usepackage[round]{natbib}
	
\begin{document}

\title{Follow the Leader: Technical and Inspirational Leadership in Open Source Software}

\bigskip 

\author{
Jerome Hergueux\thanks{ French National Center For Scientific Research (CNRS, BETA lab), Strasbourg, France and ETH Zurich, Center for Law and Economics, Zurich, Switzerland. Correspondence: \texttt{jerome.hergueux@gess.ethz.ch} } \,
Samuel Kessler\thanks{ University of Oxford, Oxford, UK. Correspondence: \texttt{samuel.kessler@robots.ox.ac.uk}} 
}

\maketitle

\begin{abstract}

We conduct the first comprehensive study of the behavioral factors which predict leader emergence within open source software (OSS) virtual teams. We leverage the full history of developers' interactions with their teammates and projects at \texttt{github.com} between January 2010 and April 2017 (representing about 133 million interactions) to establish that -- contrary to a common narrative describing open source as a pure ``technical meritocracy'' -- developers' communication abilities and community building skills are significant predictors of whether they emerge as team leaders. Inspirational communication therefore appears as central to the process of leader emergence in virtual teams, even in a setting like OSS, where technical contributions have often been conceptualized as the sole pathway to gaining community recognition. Those results should be of interest to researchers and practitioners theorizing about OSS in particular and, more generally, leadership in geographically dispersed virtual teams, as well as to online community managers.  

\bigskip

\noindent \textbf{Keywords:} Open Source Software; Leadership; Communication; Virtual Teams.

\bigskip

\noindent \textbf{To cite this paper :} 

\noindent Hergueux, Jerome and Samuel Kessler (2022). Follow the Leader: Technical and Inspirational Leadership in Open Source Software, \textit{Proceedings of the ACM CHI Conference on Human Factors in Computing Systems 2022}, ACM, New York, NY, USA. 

\noindent \href{https://doi.org/10.1145/3491102.3517516}{https://doi.org/10.1145/3491102.3517516}

\medskip

\end{abstract}

\clearpage

\section{Introduction}
Open source software (OSS) represents one of the most significant organizational innovation of the 20th century~\citep{benkler2006wealth}, where millions of developers from around the world voluntarily self-organize in virtual teams and coordinate successfully in the absence of price signals and without any pre-specified design rule or formal leadership~\citep{crowston2012free,faraj2011knowledge,levine2013open}. OSS is responsible for most of the basic utilities on which the Internet runs (e.g., the Apache web server), popular programming languages (e.g., Python) and programming environments (e.g., Eclipse). It also competes with many of its proprietary counterparts in the realm of end-user applications (e.g., Android), and operating systems (e.g., Linux). As of today, most businesses and public organizations rely on OSS for their daily activities ~\citep{walli2005growth,ghosh2007economic,greenstein2014digital}. 

Traditionally, open source software has been described as a ``technical meritocracy,'' where ``code is king''~\citep{scacchi2007free}. Developers acquire and retain influence over their followers -- that is, become community leaders -- by writing elegant code that ``just works'' ~\citep{raymond1999cathedral,weber2004success,marlow2013impression}. This view resonates with economists' theoretical conceptualization of the issue of leader emergence in work environments such as open source, where the leader's task is to get individuals to follow him \textit{voluntarily}. The problem in this case is that developers cannot credibly commit to communicating what they think is the real value of their projects. All have an incentive to claim that the community's time will be most productively spent on their own ideas. Communication is ``cheap'', and therefore ignored. The only way to solve this information asymmetry problem is to ``lead by example''~\citep{hermalin1998toward,hermalin2012leadership}, that is, exert costly effort in order to credibly signal one's belief that a project really is worth spending time on. As a result, open source leaders are simply predicted to be those who are willing to exert such effort by writing more and higher quality code than their peers. 

In sharp contrast to this traditional view, others in the open source software community insist on communication as the real key to leader emergence ~\citep{fogel2005producing,bacon2012art}. This view, in turn, resonates strongly with the management literature, which emphasizes that in order to get others to follow, leaders needs to be ``inspirational.'' The theory of inspirational leadership predicts that the individuals most likely to emerge as community leaders are those who are able to communicate effectively along several important dimensions. First, the leader needs to articulate a \textit{compelling vision} for the future. By doing so, they shape the beliefs of their followers to reduce strategic uncertainty and avoid coordination failures ~\citep{brandts2007s,foss2009towards,brandts2014legitimacy}. Second, they need to provide a \textit{common identity} to their followers, effectively shaping their preferences to get them to internalize the leader's collective goals as their own \citep{elsbach1996members,gioia1996identity,bass1999two}. Third, the leader needs to appropriately \textit{leverage emotions} to facilitate the internalization of their message \citep{forgas1995mood,bolte2003emotion,topolinski2009architecture}, and generate positive affective bounds among team members so as to maintain trust and cooperation \citep{trice1993cultures,gardner1998charismatic,jones1998experience,barling2000transformational}. The use of positive emotions is expected to be unambiguously effective in the context of this literature, but negative emotions also have a role to play in the leader's discourse, as they can foster critical thinking when most needed and help people coordinate in the face of a common threat ~\citep{george2000emotions}. Finally, the leader has to appear \textit{confident and resolute}, so that followers trust that they will stick to their ``vision'' in the long-run, while retaining an ability to adjust in the face of changing circumstances ~\citep{bolton2008leadership,vidal2007should}.   

Insofar as good community leaders are able to attract and motivate followers, the issue of leader emergence in peer production teams such as open source emerges as a high-stake research topic not only for researchers interested in theorizing about OSS or, more generally, geographically dispersed virtual teams, but also for platform designers. If the traditional view holds, then design efforts should continue to focus on improving development tools such as, e.g., version control systems, which allow developers to get a quick and reliable sense of the quantity and quality of individual code contributions. If, however, developers further reveal a preference for inspirational communication, investments may be made into developing and testing features that facilitate socialization and communication. This is also an important practical issue for project creators and community managers, as most open source projects fail to attract, and struggle to maintain, a significant contributor base ~\citep{crowston2012free}. Surprisingly enough, however, very little research to date has focused on the issue of leader emergence in open source software. 

In this paper, we leverage the full record of developers' interactions with their teammates and projects at \texttt{github.com} between January 2010 and April 2017 (i.e., 7 years and 3 months) to fill this gap in the literature. We operationalize the leadership construct based on developers' activity traces across development teams, and define variables to capture the main components of the technical meritocracy and inspirational leadership theories. We then use those variables in a survival analysis framework to identify and contrast the behavioral characteristics that are most strongly associated with leader emergence in OSS virtual teams.

\section{Relationship to previous literature}

Existing research on leader emergence in OSS virtual teams is relatively scarce. Seminal papers have focused their analysis on the largest, most successful OSS projects such as GNOME and Debian  ~\citep{o2007emergence,dahlander2011progressing}. Those projects are convenient to study since, unlike the overwhelming majority of projects which are typically organized around a small team of contributors, they tend to define leadership roles in a formal way. The main conclusion of this literature is that emergent leaders can be characterized as those who are trusted with ``lateral authority over project tasks'', and ``spend significant time coordinating project work'' ~\citep{dahlander2011progressing}. 

In general, very little research focuses on more representative samples of smaller scale OSS virtual teams ~\citep{crowston2012free}. As far as OSS leadership is concerned, two exceptions stand out. First, ~\cite{giuri2008explaining} leverages data from \texttt{sourceforge.net} to conduct a cross-sectional study of whether the diversity of developers' skills explains the likelihood that they are identified as a ``project manager'' on at least one of the projects to which they contribute. The analysis relies on the fact that SourceForge developers have the possibility to publicly report their experience in 33 types of skills, grouped into three categories: programming languages (e.g., C++, Python), application-specific skills (e.g., networking, security) and spoken languages. They find that project leaders have a relatively more diversified skill set, consistent with the idea that they are trusted to ``select the inputs provided by various participants'' and ``coordinate their efforts'' \citep{giuri2008explaining}.  

Second, ~\cite{li2012leadership} conducts a survey among 118 OSS developers registered with SourceForge to estimate the impact of leadership style on motivation and effort, as measured by the number of hours developers report working on the project per week. They find that OSS developers who describe their project leaders as trying to stimulate and inspire them declare being more intrinsically motivated and working longer hours. 

We contribute to this line of research in several ways. First, we build upon previous findings on the nature of leadership in open source virtual teams to define the construct based on developers' field activity (i.e., their project interactions). This approach allows us to operationalize leadership empirically across the universe of OSS projects hosted at \texttt{github.com}. Hence we can perform the first comprehensive investigation of the antecedents of leader emergence within OSS virtual teams. 

Second, we set-up our study so as to explicitly contrast the empirical relevance of the technical meritocracy and inspirational leadership views. By doing so, our goal is to push the empirical literature beyond its current focus on the role of technical skills in leader emergence. Although not necessarily contradictory, both views have typically been discussed separately ~\citep{scacchi2007free,raymond1999cathedral,weber2004success,marlow2013impression,fogel2005producing,bacon2012art}. Both connect to well established but distinct theoretical traditions within the fields of economics (``lead by example'') ~\citep{hermalin1998toward} and management (``lead by inspiring'') ~\citep{zehnder2017productive}. We therefore think that connecting them provides a useful framework to organize the conversation around leader emergence in OSS and, more generally, geographically dispersed virtual teams. 

Finally, we rely on the rich body of work on inspirational leadership cited in the introduction to formulate hypotheses as to how inspirational leaders are expected to communicate. Indeed, a finding that emergent leaders communicate more is not enough to support the inspirational leadership view. If, as previous research suggests, one core leadership task is to engage in coordination work, emergent leaders will be expected to communicate significantly more -- even in a pure technical meritocracy world. Effectively testing the inspirational leadership view therefore requires that researchers start to look at \textit{how} emergent leaders express themselves. 

\section{Data and variables}
\label{sec:data_vars}
We obtain our data from the GHTorrent project \citep{gousios2012ghtorrent}, which provides a queriable mirror of the data available from the Github API. The project documents all Github events occurring on public OSS projects. We collect developers' activity data at a monthly frequency over 87 consecutive months (i.e., from January 2010 to April 2017). We include in our sample all projects that have at least 6 contributors, defined as developers who have committed to the main repository at least once over the entire time period, or raised a pull request which got merged. Indeed, at a theoretical level, the concept of leadership  -- defined both in terms of peer recognition and engagement in team coordination work -- appears most relevant when teams include some minimal number of developers.\footnote{Another reason for this design choice is technical: including the high number of Github projects which only attract a handful of contributors over the course of their history leads our server (set-up with 250GB of RAM) to run out of memory.} In Appendix \ref{q4results}, we explore the generalizability of our results to teams of various sizes by replicating our main analysis separately for each quartile of team size in our data. Organization accounts, which are typically used by businesses and whose projects may not be community developed \citep{tsay2014let}, are excluded from the sample, together with fake users (this is an internal flag in GHTorrent), and we ignore forked projects.\footnote{Note further that beyond software development, Github is sometimes used by individuals for other purposes, such as website hosting. The fact that we focus on projects that receive pull requests and/or commits from at least 6 different developers alleviates this potential source of noise in our data, however.} We end-up with a final sample of 2,011,159 developers contributing to 1,951,528 distinct projects over the time period we consider (i.e., 7 years and 3 months). Figure \ref{fig:data_eng} provides a visual representation of our data engineering workflow.

\subsection{Dependent variable} 

Consistent with previous work on leadership in open source ~\citep{o2007emergence,dahlander2011progressing} and other peer production communities ~\citep{johnson2015emergence}, we define a leader as a developer who 1) is a \textit{recognized}, \textit{trusted} and \textit{influential} participant in a given OSS project, and 2) engages in significant \textit{coordination work} among team members. We derive 3 alternative criteria to capture those constructs in the Github context. 

In Github, the primary way open source projects attract new code contributions is through pull requests ~\citep{marlow2013impression}. The potential contributor forks the source code to implement the desired modifications on their local machine, and then raises a request that those changes be ``pulled'' into the main repository. Arguably, the clearest indication that a developer is recognized by their peers within a given OSS team is for them to have write access to the source code and be engaged in evaluating the pull requests submitted by others for suitability. Developers who engage in such reviewing tasks have authority to decide which pieces of code deserve to get in. They are both trusted and influential members of their community ~\citep{tsay2014let,gousios2014exploratory}. This is one criterion we use to define team leadership in our setting. 

\begin{figure}
     \centering
     \begin{minipage}{.54\textwidth}
     \centering
     \includegraphics[width=0.9\linewidth]{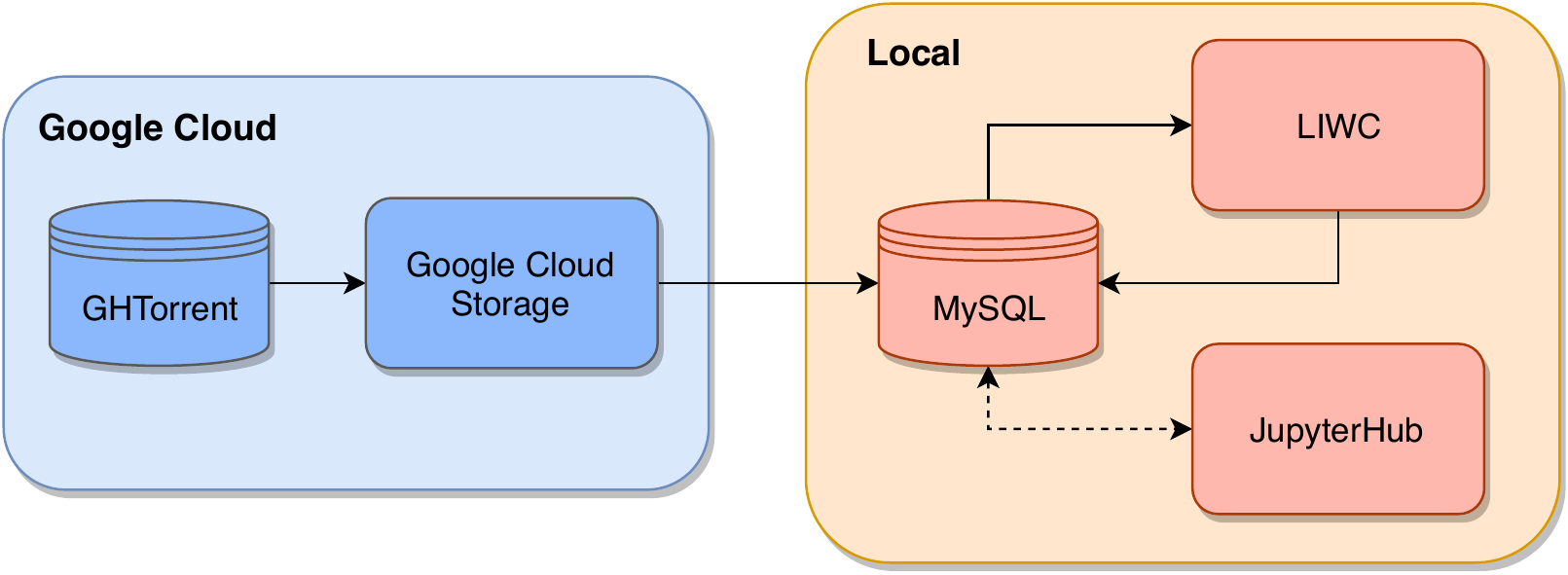}
     \caption{Overview of our data engineering workflow. The GHTorrent database is accessible on Google Cloud and can be queried with Google BigQuery. The events data is exported as csv files and organized in a local MySQL database. The text of developers' comments is exported as json files and uploaded to the LIWC software. The relevant communication features are computed and uploaded back to our local database. The solid arrows represent the flow of data, and the dashed arrows represent interactions with the data.}
     \label{fig:data_eng}
     \end{minipage}
     \hfill
     \begin{minipage}{.42\textwidth}
         \vspace{-35.0mm}
         \centering
         \includegraphics[width=0.9\linewidth]{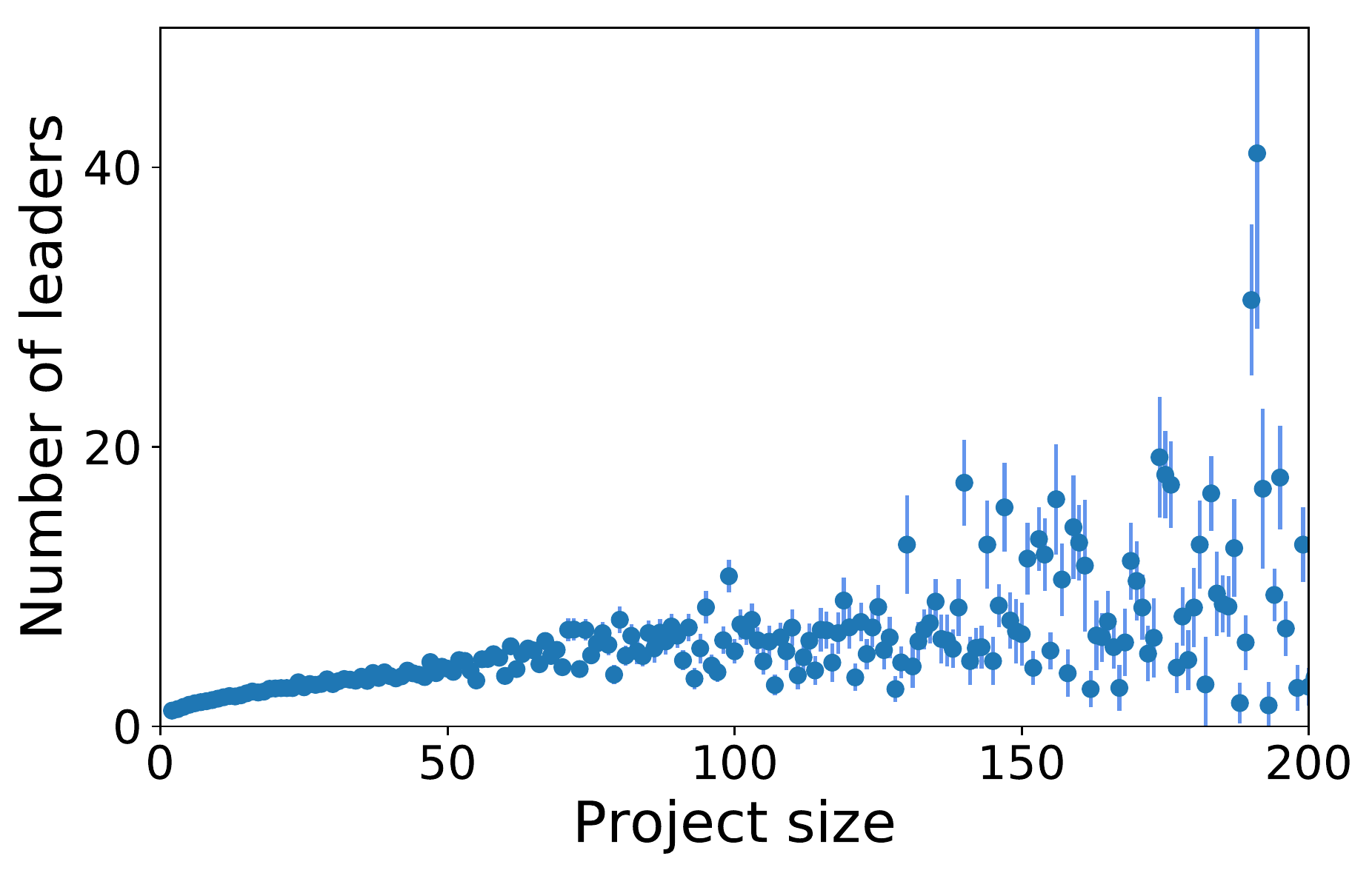}
         \caption{Average number of leaders as a function of team size (truncated if \texttt{nb developers in team} $>$ 200). The vertical lines denote the 95\% confidence interval around each mean value.}
         \label{fig:nb_leaders}
     \end{minipage}
\end{figure}

The second and third criteria relate more directly to leaders' coordination role within their respective teams. Specifically, instead of reviewing a pull request himself, a developer with commit rights can decide to assign it to another developer for evaluation (perhaps because they have some expertise on the subject matter). The same principle holds for the issues that are raised with respect to the project, which can also be assigned by one developer to another to solve. We consider those events to be indicative of the fact that a developer is recognized by their peers and trusted to engage in significant coordination work. 

All in all, we consider that a developer emerges as a leader within their team at the earliest date at which we observe him performing one of the following actions:     

\begin{enumerate}
    \item merge the pull request of another developer in the main repository;
    \item assign a pull request for review to another developer;
    \item assign an issue to solve to another developer.
\end{enumerate}

Figure \ref{fig:nb_leaders} reports the average number of leaders by team size (together with their 95\% confidence intervals). We can see that the number of leaders increases as development teams grow, but less than proportionally. On average, bigger teams therefore have a smaller proportion of their contributors who behave as team leaders. The estimates become quite imprecise at the far right of the graph, due to the relatively small number of very large Github development teams.

\subsection{Explanatory variables} 

The explanatory variables which we track capture developer behavior associated with leader emergence, as theorized by the technical meritocracy and inspirational leadership views. As stated in the introduction, the technical meritocracy view predicts that emerging open source leaders are those who ``lead by example'', that is 1) are \textit{more committed} to the project than their peers (i.e., participate more), and 2) write \textit{higher quality code} on average. We capture those constructs through several variables, all of which are documented at the developer $\times$ project $\times$ month level:

\begin{enumerate}
    \item Measures of project commitment:
        \begin{itemize}
            \item \textbf{nb pull requests (PR)}: the monthly number of pull requests raised for consideration by a developer in a project ~\citep{tsay2014influence};
            \item \textbf{nb issues raised}: the monthly number of project issues raised by the developer;
            \item \textbf{nb messages on PR \& issues}: the monthly number of messages sent by the developer on raised issues and pull requests. 
        \end{itemize}

    \item Measures of code quality:
        \begin{itemize}
            \item \textbf{nb updates on PR}: the monthly number of updates made by the developer on their own pull requests (a higher number of updates indicates that the code was of lower initial quality);
            \item \textbf{nb inline code comments}: the monthly number of inline code comments received by the developer on their pull requests. Inline code comments can be used by reviewers either to document running code or to pinpoint issues and potential improvements ~\citep{kalliamvakou2014promises,gousios2014exploratory,tsay2014let}. The sign of the aggregate relationship between this variable and leadership emergence prospects is therefore ambiguous \textit{a priori}.
        \end{itemize}
\end{enumerate}

Note that, like all the other variables in this paper, the above measures of code quality are based on the data available from GHTorrent on developers' interactions with their own project code. This approach to code quality as measured through developers' digital activity traces contrasts with an ``external'' evaluation approach where the source code of each project would have to be downloaded directly from Github for processing. We follow the former strategy because it does not require us to operationalize code quality across the universe of developers, software projects and programming languages supported by Github, but rather infers code quality from the actions of developers themselves. While this approach is not without limits, it has the benefit of generalizability. As of today, the software engineering literature has not yet converged on a scalable set of methods to capture the multifaceted aspects of code quality (e.g., conciseness, portability, maintainability, testability, reliability, structuredness, efficiency...) across programming languages and coding practices. (For more on this active research topic, see  \cite{pantiuchina2018improving}, \cite{borstler2018know} , \cite{molnar2019longitudinal}, \cite{sharma2021survey}.)

On the other hand, the inspirational leadership view predicts that emerging open source leaders are those who are able to communicate effectively in order to 1) articulate a \textit{compelling vision} for the future of the team, 2) unite followers under a \textit{common identity}, 3) \textit{leverage emotions} to facilitate the internalization of their message and maintain trust-based relationships among team members, and 4) \textit{appear confident and resolute} enough that followers trust that they will stick to their stated ``vision.'' 

In order to capture those 4 constructs, we concatenate the messages posted by each developer on all issues and pull requests at the project $\times$ month level. We then pass this text through a Natural Language Processing software, the Linguistic Inquiry and Word Count (LIWC) ~\citep{tausczik2010psychological}, to generate measures aimed at operationalizing each.\footnote{In order to maximize the reliability of the scores we compute, this step requires that the monthly set of messages posted at the developer $\times$ project level is at least 250 words long (see the best practices reported in ~\cite{tausczik2010psychological}).} LIWC counts the occurrences of items in the raw text that match a pre-specified dictionary tailored to generate psychologically insightful categories. For instance, the \textit{focusfuture} dictionary contains 97 items such as ``will'', or ``soon''. The \textit{posemo} dictionary contains 620 items such as ``love'', ``nice'', or ``sweet''. The \textit{negemo} dictionary contains 744 items such as ``hurt'', ``ugly'', or ``nasty''. The \textit{certainty} dictionary contains 113 items such as ``always'' or ``never''. Finally, the \textit{tentative} dictionary contains 178 items such as ``maybe'' or ``perhaps''.\footnote{While the full dictionaries are not made publicly available, more details on their development can be found here: \href{https://repositories.lib.utexas.edu/bitstream/handle/2152/31333/LIWC2015_LanguageManual.pdf}{https://repositories.lib.utexas.edu/bitstream/handle/2152/31333/LIWC2015\_LanguageManual.pdf}} 

Previous research has found LIWC to perform well in electronic messaging contexts such as Twitter ~\citep{golder2011diurnal,kivran2011network}, microblogging ~\citep{de2013understanding}, Wikipedia ~\citep{iosub2014emotions} and Github \citep{pletea2014security}. This paper extends this line of work and uses LIWC to operationalize inspirational communication in OSS developers' written exchanges. Overall, we recover 6 communication variables from LIWC, documented at the developer $\times$ project $\times$ month level:

\begin{enumerate}
    \item Articulate a forward thinking narrative, captured by:
        \begin{itemize}
            \item \textbf{prop(focusfuture)}: the proportion of words in text that are forward looking.
        \end{itemize}
    \item Unite followers under a common identity, captured by: 
        \begin{itemize}
            \item \textbf{prop(we)}: the proportion of first person plurals (e.g., \emph{``we''}, \emph{``us''}, \emph{``our''}) in text.
        \end{itemize}
    \item Leverage emotions, captured by: 
        \begin{itemize}
            \item \textbf{prop(posemo)}: the proportion of words in text that express positive emotions,
            \item \textbf{prop(negemo)}: the proportion of words in text that express negative emotions.
        \end{itemize}
    \item Appear confident and resolute, captured by:
        \begin{itemize}
            \item \textbf{prop(certain)}: the proportion of words in text that express certainty,
            \item \textbf{prop(tentative)}: the proportion of words in text that express tentativeness.
        \end{itemize}
\end{enumerate}

In Table~\ref{tab:liwc_examples}, we show comments from the top percentile of the distribution for each inspirational communication variable and illustrate how LIWC allows to identify developers' communication style. In addition we show in Appendix~\ref{vader} that our LIWC scores match up to more sophisticated methods such as NLTK VADER, while being much less computationally intensive to run. Future research might nonetheless explore the relevance of a different approach to this measurement problem: train a dedicated machine learning algorithm on human-rated data along each dimension. This would have the benefit of specificity, but is not without costs and challenges (see, e.g., \cite{lin2020survey} or \cite{ram2018supervised} for a discussion in the context of sentiment analysis).

Based on the above cited management literature on inspirational leadership, we hypothesize the variables \textit{prop(focusfuture)}, \textit{prop(we)}, \textit{prop(posemo)} and \textit{prop(certain)} to be positively associated with leader emergence. Conversely, we hypothesize \textit{prop(tentative)} to be negatively associated with leader emergence prospects. The impact of \textit{prop(negemo)} is theoretically ambiguous. Negativity runs contrary to most communication strategies aimed at fostering the internalization of a leader's message and promoting cooperation among followers. (Those strategies typically rely on positive affect.) However, negative emotions are powerful tools to foster critical thinking and unite a group when it faces challenges ~\citep{george2000emotions}. 

Finally, note that beyond hypothesis testing, one important goal of our empirical investigation is to be able to directly compare the magnitudes of the coefficients on our technical and inspirational leadership variables. This is made difficult by the fact that all variables are expressed in different units which are not easy to interpret. In order to get around this issue, we standardize each explanatory variable $x$ as $z = {x - \bar{x} \over s}$, where $\bar{x}$ is the sample mean and $s$ is the sample standard deviation of the variable considered. This linear transformation does not affect statistical inference, but it allows to interpret each coefficient as the effect of moving ``one standard deviation away'' from the sample mean for the variable considered -- effect sizes can now be directly compared.

\subsection{Control variables}

As explained above, extant research on leadership in open source has identified coordination work -- and therefore \textit{communication} -- as one of the defining tasks performed by project leaders ~\citep{o2007emergence,dahlander2011progressing}. Since we follow this literature to operationalize leadership in our setting, we expect developers who become project leaders to communicate significantly more than their fellow team members. Leaders' relative focus on coordination work does not necessarily imply inspirational leadership, however. Testing the inspirational leadership view therefore requires that one looks at \textit{how} leaders express themselves for a given volume of communication. Beyond our explanatory variables of interest, our analysis therefore controls for the total size of the text posted by project developers in any given month with the variable \textit{log(1 + word count)}. This control variable should be strongly associated with leader emergence -- although not necessarily in a causal way.

Second, we expect the total number of developers in a team to have an independent impact on leader emergence prospects, e.g., because larger projects have more mature processes in place which make it more difficult for new developers to emerge as team leaders. This hypothesis seems supported by Figure \ref{fig:nb_leaders}, where we can see that larger projects tend to be led by a smaller proportion of developers. We therefore include the variable \textit{log(nb developers in team)} as a control in our analysis. 

Third, we rely on Github's ``star'' feature, which developers use to signal their interest in a given project, to compute a measure of project popularity ~\citep{dabbish2012social,tsay2014influence,klug2016understanding}. We expect it to be more challenging for developers to emerge as team leaders on projects which are highly popular within the extended open source community. This measure, computed as \textit{log(1 + nb project stars)}, could be highly correlated with \textit{log (nb developers in team)} ~\citep{tsay2014influence}, but since we have a lot of degrees of freedom in our model, we do not worry about statistical power \citep{goldberger1991course}.\footnote{See Appendix \ref{pwcorr} for a detailed analysis of the pairwise correlation structure of our variables.}

Last, we expect a developer who already benefits from a favorable reputation in the extended open source community to have an easier time emerging as a leader within some new team. We capture this through the variable \textit{log(1 + developer reputation)}, with \textit{developer reputation} defined as the cumulative sum of the number of stars received by all the projects on which the developer is already a leader by any given month\footnote{We add 1 to this number to ensure that the developer still gets a positive score if the projects they are a leader on have no stars, effectively distinguishing them from non-leaders in the data (i.e., $\log(1+1) > \log(1) = 0$).}. More precisely, if $P$ is the set of projects on which the developer is a leader at some time $t$, and $S_{t, j}$ is the number of stars received by project $j$ in month $t$, then \textit{log(1 + developer reputation)} is defined with:

\smallskip
\begin{equation}
\label{equation::exponential_weights_update}
    \textit{developer reputation} = \begin{cases}
        0 \quad &\text{if } P = \emptyset \\
        1 + \sum_{t'\leq t} \sum_{j \in P} S_{t', j} &\text{if } P \neq \emptyset.
        \end{cases}
\end{equation}
\smallskip

\begin{table*}
\caption{Descriptive statistics for all covariates. Regarding the summary statistics construction, all \textit{Technical Leadership}, \textit{Inspirational Leadership} and \textit{log(1 + word count)} variables are indexed by developer $\times$ project key pairs, so grouped at this level and averaged over time. The \textit{log(nb developers in team)} and \textit{log(1 + nb project stars)} variables are indexed by project, thus grouped by project and averaged over time. Finally the \textit{log(1 + developer reputation)} variable is indexed at the developer level and so grouped by developer and averaged over time. The time period considered is from January 2010 to April 2017, i.e. 87 monthly time points.} 
\centering
\begin{tabular}{lccccc} 
 \toprule
  \bf{Variable} & \bf{min} & \bf{mean} & \bf{median} & \bf{max} & \bf{std} \\
 \midrule
 \textit{Technical Leadership} \\
{ } \texttt{nb pull requests (PR)} & 0.000 & 0.005 & 0.000 & 86.000 & 0.102 \\
{ } \texttt{nb issues raised} & 0.000 & 0.086 & 0.000 & 4,268.000 & 3.343 \\
{ } \texttt{nb messages (PR \& issues)} & 0.000 & 0.107 & 0.000 & 240.938 & 0.690 \\
{ } \texttt{nb updates on PR} & 0.000 & 0.041 & 0.000 & 252.535 & 0.358 \\
{ } \texttt{nb inline code comments} & 0.000 & 0.061 & 0.000 & 146.000 & 0.603 \\
\textit{Inspirational Leadership} \\
{ } \texttt{prop(focusfuture)} & 0.000 & 0.002 & 0.000 & 5.140 & 0.036 \\
{ } \texttt{prop(we)} & 0.000 & 0.002 & 0.000 & 7.190 & 0.038 \\
{ } \texttt{prop(posemo)} & 0.000 & 0.005 & 0.000 & 49.645 & 0.091 \\
{ } \texttt{prop(negemo)} & 0.000 & 0.002 & 0.000 & 6.730 & 0.030 \\
{ } \texttt{prop(certain)} & 0.000 & 0.002 & 0.000 & 6.760 & 0.042 \\
{ } \texttt{prop(tentative)} & 0.000 & 0.005 & 0.000 & 17.330 & 0.089 \\
\textit{Controls} \\
{ } \texttt{log(1 + word count)} & 0.000 & 0.014 & 0.000 & 11.359 & 0.226 \\
{ } \texttt{log(nb developers in team)} & 1.792 & 2.442 & 2.197 & 9.993 & 0.754 \\
{ } \texttt{log(1 + nb project stars)} & 0.000 & 1.373 & 0.693 & 10.588 & 1.687 \\
{ } \texttt{log(1 + developer reputation)} & 0.000 & 0.204 & 0.000 & 11.054 & 0.850 \\
\bottomrule
\end{tabular}
\label{tab:explanatory_variables}
\end{table*}

\begin{table}
  \centering
  \footnotesize
  \caption{\normalsize{Examples of developer messages with high inspirational communication LIWC scores. For each variable, the messages were extracted from the top percentile of developers' monthly project communication scores.}} \label{tab:liwc_examples}
    \scalebox{0.65}{
    \begin{tabular}{l}
    \toprule
    \toprule
    {\normalsize \textbf{focusfuture :} } \\
    \midrule
    { } - ``Closed for a while. We will discuss about this in the planning of release 0.3.'' \\
    { } - ``We're very unlikely to address the /watch component of this feature in the foreseeable future. A new issue will be filed if this changes; \\ { } file a new issue if you feel strongly that this should be reconsidered.'' \\
    { } - ``Once tests pass, I'm going to go ahead and merge this myself because I'm the only FTE on this project. Going forward, I'm going to be \\ { } merging my own pull requests on this project because your schedule isn't predictable. Please continue to assign pull requests to me.'' \\
    { } - ``Sorry for the late response. I'll be helping maintain this plugin for the foreseeable future. If any of these issues are still persisting (all of the \\ { } tickets are quite old), please let me know and I will do what I can to try and help fix things. This should be fixed with the latest version that \\ { } I merged in. Please let me know.'' \\
    \midrule
    \normalsize{\textbf{we :}} \\
    \midrule
    { } - ``We don't like to use these kind of smart pointer, we think we can control raw pointer well. We are the guys who think c++ is c with class. \\ { } So we don't use lots of c++11, c++14 feature.'' \\
    { } - ``We've got a lot of developers now, we need to make sure we're all writing code in the same way or we'll be adding barriers to working \\ { } well as a team!'' \\
    { } - ``Okay, give us the earliest date that we can do it, and we'll take care of it. We were approached with this before and made all the changes \\ { } only to discover that we were too hasty and needed to go back and revert.'' \\
    { } - ``Thanks for your Pull Request, we really appreciate people trying to improve our first open source library :) 2015 is an intense year for us \\ { } regarding our workload. We are planning to, finally, upgrade `scredis` to support the latest Redis features. Once we jump on the task, we'll \\ { } make sure to take a look at your changes as they make sense. As we want to do things right, we can't merge it right away. Thanks for your \\ { } understanding :)'' \\
    \midrule
    \normalsize{\textbf{posemo :}} \\
    \midrule
    { } - ``Looks good, thanks! :+1:'' \\
    { } - ``Wow! That's really cool. Thanks for continuously working on this, I really really appreciate it. :)'' \\
    { } - ``Thank you so much for communicating with me through your process and it looks absolutely lovely! I really appreciate your contribution!'' \\
    { } - ``This is fantastic! It is such a cute and creative addition to the website. I'm super grateful for your contribution. Thank you :)'' \\
     \midrule
    \normalsize{\textbf{negemo :}} \\
    \midrule
    { } - ``You got to be kidding. Me? Author...hell no, its an obvious mistake !'' \\
    { } - ``This change is wrong.'' \\
    { } - ``:angry: :angry: :angry: :angry: :angry: :angry: :angry:'' \\
    { } - ``IT'S THE RETURN OF THAT ONE OBSCURE BUG AGAIN AAAAAAAAAAAAAAAAAAAAAAAAAAAAAAAAAAAAAAAAAAAAAA \\ { }AAAAAAAAAAAAAAAAAAAAAAAAAAAAAAAAAAAAAAAAAAAA'' \\
     \midrule
    \normalsize{\textbf{certain :} }\\
    \midrule
    { } - ``Can you be more specific? I just tried it with the auto-generated settings file and that works fine, as expected. All definitions are true, and \\ { } the default theme is true too.'' \\
    { } - ``The error message is obscure, I'll grant, =), but the docstring does specify that spacing must be length 3! We don't currently support \\ { } spacing in 2D images because it's rare to have different voxel spacing in just 2D images. (I have never come across this.) Do you have a \\ { } use-case for 2D spacing? If so, I can add support for that. If not, I'll add a `ValueError` that explains what went wrong better than the current \\ { } message. Thanks for the report!'' \\
    { } - ``I just pulled this a few seconds ago and it worked just fine.'' \\
    { } - ``That is to create the branch. But we will have to switch to it by `git checkout response`. A short cut if you haven't seen it is `git checkout -b \\ { } response`. This will create the branch and check it out.'' \\
    \midrule
    \normalsize{\textbf{tentative :} }\\
    \midrule
    { } - ``Is there a reason we'd want the debug version? Maybe this should be a constant at the top of the file if so, or something.'' \\
    { } - ``Seems like we can get a vendor id but have to translate it to a string ourselves. Probably we can either just use the numbers or have a few \\ { } special numbers we translate to strings. Seems like we can't get renderer, but I'm not sure? Would be useful.'' \\
    { } - ``I notice this is only checked for subdata, should it check for setdata too? Or is an error only possible with a > 0 offset? I guess if it doesn't \\ { } set the data, it should probably log?'' \\
    { } - ``To be honest it's because I don't need any of the additional formatting options provided by pg-promise, and when in doubt I try not to use \\ { } features I don't need. Plus I (maybe wrongly, i'm not sure) assumed that the node-postgres would handle it faster, if there were any bugs that \\ { } the larger library would have a better chance at getting them fixed for me, and it was unclear if i could use prepared statements (which was \\ { } a must for me).'' \\
    \bottomrule
    \bottomrule
    \end{tabular}
    }
    \vspace{1.0cm}
\end{table}

\subsection{Dataset overview}

Table~\ref{tab:explanatory_variables} provides some summary statistics on all the right hand side variables used in our model. We organize our data as a panel at the developer $\times$ project $\times$ month level, and cover a time period spanning from January 2010 up to April 2017 (i.e., 87 consecutive months). The dataset comprises 132,094,359 observations including 154,031 leader emergence events. Our explanatory variables are all captured at this level of granularity, with some exceptions for our control variables. The total number of developers in a given team only varies at the project level, and is defined as the total number of developers who contributed code to the project by the end of our time period. The number of project stars is defined as a cumulative sum and therefore varies at the project $\times$ month level. Developer reputation is also defined as a cumulative sum, and varies at the developer $\times$ month level. Note that we log-transform those control variables before including them in the model. This transformation allows to interpret the effect of those variables in terms of percentage changes as opposed to levels, which might be both easier to interpret and more realistic. For instance, instead of interpreting our coefficient on team size as the effect of adding one more developer to the final number of contributors to a given project, irrespective of whether we start from a very low or a very large number of team contributors, we now interpret it in terms of \textit{percentage change} in final team size \citep{wooldridge2010econometric}.

We describe our explanatory variables in Table~\ref{tab:explanatory_variables} by averaging them over time at the developer $\times$ project level, and then aggregating by taking a mean. Control variables are described following the same logic, but at the relevant level of granularity. The most notable feature of the dataset is that it features a large number of zeros, with all variables distributed as power laws -- a natural characteristic of participation in digital spaces such as open source, online message boards or Wikipedia. 

Finally, Table~\ref{tab:liwc_examples} illustrates our LIWC variables by providing examples of messages extracted from the top percentile of the distribution for each inspirational communication measure.

\begin{figure}
     \hspace{0.05\textwidth}
     \begin{minipage}[c] {0.54\textwidth}
     \caption{Representation of the censored OSS observational dataset. The study begins in 01/2010 and ends in 04/2017. Some developers become leaders during this time period, hence a leadership event $T$. Right censoring implies that for all other developers, we never get to observe if, and if so when, they become leaders at some later point in time.}
     \label{fig:censoring}
     \end{minipage} 
     \hfill 
     \hspace{0.05\textwidth}
     \begin{minipage}[c] {0.40\textwidth}
     \begin{tikzpicture}
     \vspace{-2.0mm}
     \draw[dashed]  (0,0) node[below] {$01/2010$} -- (0,3);
     \draw[dashed] (3,0) node[below] {$04/2017$} -- (3,3);
     \draw (0, 2.7) -- (2,2.7) node[fill,circle,inner sep=0.01cm,minimum size=0.1cm] {} node[above] {$T$};
     \draw (1, 2.2) -- (2,2.2) node[fill,circle,inner sep=0.01cm,minimum size=0.1cm] {} node[above] {$T$};
     \draw (0.5, 1.7) -- (3,1.7) node[draw,circle, fill=black, minimum size=0.1cm, inner sep=0.01cm] {};
     \draw (2.5, 1.2) -- (3,1.2) node[draw,circle, fill=black, minimum size=0.1cm, inner sep=0.01cm] {};
     \draw (1.5, 0.7) -- (3, 0.7) node[draw,circle, fill=black, minimum size=0.1cm, inner sep=0.01cm] {};
     \draw (1.2, 0.2) -- (2.7, 0.2) node[fill,circle,inner sep=0.01cm,minimum size=0.1cm] {} node[above] {$T$};
     \end{tikzpicture}
     \end{minipage}
\end{figure}

\section{Empirical strategy}
\label{empirical_strategy}

Our data is organized as a panel at the developer $\times$ project $\times$ month level. Based on our covariates, we seek to explain the probability that any given team member emerges as a leader at time $t$, with $t = 0$ for the first month the developer interacts with the project by either submitting a pull request which gets merged, or committing code to the main repository. As can be seen from Figure~\ref{fig:censoring}, the most notable feature of our data is that it is right-censored, i.e., we do not get to observe who becomes a leader in any given team after April 2017. In order to properly account for this data structure, we rely on survival analysis in order to predict the ``leader emergence'' event as a function of time and our covariates. More precisely, because our explanatory variables are time-dependent, we rely on the Extended Cox model in order to predict the probability that some developer emerges as a leader within his team in any given month. (For a thorough treatment of the subject, we refer the interested reader to ~\cite{cox1984analysis}, \cite{kalbfleisch2011statistical}, \cite{aalen2008survival}, \cite{kleinbaum2010survival}.)

The instantaneous rate at which the leadership event occurs, or hazard, is described by the Extended Cox model as:

\begin{align}
\label{eq:ext_cox}
\lambda(t| \boldsymbol{X}(t)) = \lambda_0(t)\exp \left(\boldsymbol{\beta}^\top \boldsymbol{X}(t) \right),
\end{align}

where $\lambda_0(t)$ is the \emph{baseline hazard}, which is a non parametric function of time that does not depend on our covariates. The baseline hazard can be interpreted as an intercept for the model. $\boldsymbol{X}(t)$ denotes the vector of our time-varying covariates, with $\boldsymbol{\beta}$ representing the linear coefficients which we estimate. The coefficients $\boldsymbol{\beta}$ can be estimated via partial maximum likelihood without the need to estimate $\lambda_{0}(t)$. Consider two identical developers $a$ and $b$ who only differ in the value of one covariate $X_i$ at time $t$. The instantaneous hazard ratio can then be written as:

\begin{align}
    \frac{\lambda(t|X_{ai}(t))}{\lambda(t|X_{bi}(t))} = \exp(\beta_i).
\end{align}

In this case, $\exp(\beta_i)$ can be interpreted as follows: given a one unit increase in $X_i$ at time $t$, $\exp(\beta_i)$ denotes the estimated \textit{relative} change in the hazard of the leadership event between developers $a$ and $b$ at time $t$. \textit{Survival} in this model is defined as the probability $S(t| \boldsymbol{X}(t))$ that a subject will survive (i.e., not experience the leadership event $T$) past time $t$ given covariates $\boldsymbol{X}(t)$. It can be written as:

\begin{align}
    S(t|\boldsymbol{X}(t)) = P(T \geq t | \boldsymbol{X}(t)) = \exp \left( - \int_{t}^{\infty}\lambda_0(u)\exp \left(\boldsymbol{\beta}^\top \boldsymbol{X}(u) \right)du \right).
\end{align}

Note that in our case, we are more interested in $1-S(t)$\footnote{Moving forward, we drop the conditioning on $\boldsymbol{X}(t)$ for ease of notation.}, i.e., the probability that the leadership event \textit{does} occur by time $t$. The assumption of the model that the effect of time on all $\beta_i$ is independent of $t$ can be checked by looking at whether the Schoenfeld residuals depend on the survival time ~\citep{grambsch1994proportional}. A graphical inspection shows that this is not the case. We therefore present Extended Cox regression coefficients, with robust standard errors clustered at the developer level.\footnote{Developer-level clustering accounts for the fact that residuals might be correlated within each developer, without making any assumption about the form of this autocorrelation. This does not affect the point-estimates, but allows for robust statistical inference by correcting for potential heteroskedasticity in the error term. On top of developer-level clustering, we also experiment with project-level clustering, and obtain the same results. Yet another way to correct for possible autocorrelation in the error term is to run a multi-level (or ``random intercept'') model \citep{wooldridge2010econometric}, where the baseline hazard function is modified to include cluster-specific random intercept terms, the distribution of which needs to be assumed \citep{Ripatti2000,therneau2000cox}. The \texttt{coxme} \texttt{R} package \citep{therneau2015mixed} allows to implement this model. The method remains highly computationally intensive -- the maximizer never converges on our full dataset -- but we obtain consistent results with the clustering approach and when working with smaller random sample of our data.} We use the \texttt{survival} package in \texttt{R} to perform our analysis \citep{survival-package}.

\section{Results}

We present our main regression result in section \ref{main}. We then use our model to run a few theoretically interesting leader emergence simulations and counterfactual prediction exercises in section \ref{survival}.

\subsection{Main result} \label{main}

We present our Extended Cox regression coefficients together with their associated standard errors and $p$-values in Table \ref{table:result}. We can see that our technical leadership variables are significantly associated with leader emergence. Focusing first on measures of project commitment (i.e., the intensive margin of participation), all else equal, a one standard deviation increase in the number of pull requests raised in any given month is associated with a 1.4\% increase in the instantaneous leader emergence hazard. Over time, such effects can accumulate and generate important between-developer differences in leadership emergence prospects. The coefficient is significantly smaller for the number of issues raised with the project, where a one standard deviation increase in any month is associated with a 0.3\% increase in the hazard. Finally, the effect of the number of messages posted to comment on pull requests and issues actually has a negative impact on the hazard: a one standard deviation increase in the number of messages posted in any given month reduces the hazard of leader emergence by 1.3\%, all else equal.\footnote{In Appendix \ref{q4results}, we show that this result is in fact driven by the largest Github teams, i.e., those that have more than 10 developers. In smaller teams, posting more messages is actually \textit{positively} related to leader emergence. This result can be interpreted in terms of steeply rising coordination costs as the team gets larger, so that posting more messages (for a given volume of text) becomes collectively costly on balance.}

\begin{table*}
\caption{Correlates of leader emergence. The table presents Extended Cox regression coefficients with robust standard errors clustered at the developer level. All variables are standardized apart from the control variables. This done so that the magnitude of the coefficients can be directly compared and interpreted as the effect on the hazard of moving ``one standard deviation away'' from the sample mean for the variable considered at time $t$.}
\centering
\begin{tabular}{lccc} 
 \toprule
Variable & $\exp(\hat{\beta_j})$ & robust standard error & $p$-value \\
 \midrule
\textit{Technical leadership} \\
\text{ } nb pull requests (PR) &  1.014 & 5.314e-04 & < 2e-16 \\
\text{ } nb issues raised & 1.003 & 1.874e-04 & < 2e-16 \\
\text{ } nb messages (PR \& issues) & 0.987 & 1.768e-03 & 9.89e-11 \\
\text{ } nb of updates on PR & 0.967 & 5.961e-03 & 1.53e-08 \\
\text{ } nb inline code comments & 1.021 & 6.642e-04 & < 2e-16 \\
\textit{Inspirational Leadership} \\
\text{ } prop(focusfuture) & 1.005 & 1.330e-03 & 0.000113 \\
\text{ } prop(we) &  1.005 & 7.718e-04 & 1.58e-10 \\
\text{ } prop(posemo) & 1.010 & 1.918e-03 & 1.18e-07 \\
\text{ } prop(negemo) & 1.007 & 1.031e-03 & 2.23e-11 \\
\text{ } prop(certain) & 1.008 & 1.390e-03 & 6.14e-09 \\
\text{ } prop(tentative) & 1.000 & 1.696e-03 & 0.945906\\
\textit{Controls} \\
\text{ } log(1 + word count) & 1.950 & 1.415e-02 & < 2e-16 \\
\text{ } log(nb of developers in team) & 0.707 & 1.562e-02 & < 2e-16 \\
\text{ } log(1 + nb project stars) & 1.091 & 1.356e-02 & < 2e-16 \\
\text{ } log(1 + developer reputation) & 1.089 & 4.727e-03 & < 2e-16 \\
\midrule
nb observations = 132,094,359  \\
nb events = 154,031 \\
\bottomrule
\end{tabular}
\label{table:result}
\end{table*}

As expected, measures of code quality are very strongly associated with leader emergence. A one standard deviation increase in the number of updates made to one's own pull requests in any month decreases the hazard of emerging as a leader by 3.3\%. Controlling for the number of downstream code updates, the number of inline code comments received is positively associated with leader emergence: a one standard deviation increase in the number of inline comments on one's pull requests on a given month is associated with a 2.1\% increase in the hazard. Since the variable captures the effect of receiving more comments on one's code once the number of updates that those comments may have generated is explicitly controlled for, the coefficient might be simply picking up the effect of those comments which are not related to technical flaws in the source code.

Turning our attention to the inspirational leadership variables, we can see that emerging leaders do seem to employ inspirational communication strategies. A one standard deviation increase in the proportion of forward looking words at time $t$ is associated with a 0.5\% increase in the instantaneous hazard of leader emergence. The same effect size is estimated with respect to the proportion of first person plurals (e.g., ``we'', ``us'') in the text. 

Leveraging emotions in one's communication appears to have an even bigger effect on leader emergence prospects. A one standard deviation increase in the use of positive and negative emotions in any month is associated with an increase in the hazard of 1\% and 0.7\%, respectively. This suggests that naturally emerging leaders within open source software virtual teams effectively use \textit{both} positive and negative affect in their communication strategies. Note that LIWC has sometimes been criticized for its crude measurement of sentiments in Internet-mediated text, as it does not explicitly take emoticons, punctuation, or slang into account. We demonstrate the robustness of our results to using another tool specifically attuned to measuring sentiments expressed via social media -- the NLTK VADER method \citep{hutto2014vader} -- in Appendix \ref{vader}. 

Last, we find that a one standard deviation increase in the proportion of words that express certainty at time $t$ is associated with a 0.8\% increase in the leadership hazard. The effect of tentativeness, on the other hand, is precisely estimated to be zero. This is inconsistent with a strong statement of the confidence and resoluteness hypothesis, whereby expressing doubts would always be a sign of poor leadership. In this respect, our results are in line with the theoretical literature, which stresses that an efficient inspirational leader needs to appear resolute and confident enough that followers invest in their vision without fears that it might erratically change \textit{while} retaining their ability to amend it (i.e., express doubts) in the face of new information ~\citep{bolton2008leadership,vidal2007should}. 

Finally, our control variables are also related to leader emergence in interesting ways. As expected given the fact that we identify leadership with coordination work ~\citep{yoo2004emergent,johnson2015emergence}, we find developers' sheer amount of communication to be most strongly associated with leader emergence., with each percentage increase in the quantity of text posted being associated with a 95\% increase in the leadership hazard. All else equal, project-level characteristics are also significantly related to leadership emergence prospects. Specifically, in agreement with the pattern observed in Figure~\ref{fig:nb_leaders}, we find that each percentage increase in final team size reduces the instantaneous leadership hazard by almost 30\%. This result is consistent with the idea that large projects are both more attractive and successful, which makes the leadership emergence process more competitive for all developers. Surprisingly, controlling for project size, it is easier to emerge as a team leader on projects that receive a higher number of stars from the extended open source community. This result probably reflects the fact that for a team of a given size, appointing more leaders generates network effects that cause a relative increase in the number of project stars received.\footnote{This interpretation is supported by the fact that, if we exclude team size from the model, we recover a negative coefficient on this variable. Indeed, the correlation coefficient between both variables is as high as 0.5 (see Figure \ref{fig:correlation_matrix} in Appendix \ref{pwcorr}).} Last, developers' preexisting reputation within the community has a significant impact on the prospects of leader emergence at time $t$, with each percentage increase in developer reputation increasing the hazard by 8.9\%. 

All in all, the magnitudes of the coefficients we estimate suggest that, above and beyond the quantity and quality of developers' technical contributions, inspirational communication is a significant predictor of leader emergence within OSS virtual teams. Indeed, according to our model, a one standard deviation increase in all inspirational leadership variables in any given month would be associated with a 3.5\% increase in the hazard of leader emergence. By comparison, a one standard deviation increase in the number of pull requests raised, together with an associated one standard deviation decrease in the number of updates made to them (representing an increase in both the quantity and quality of code contributions at time $t$), would be associated with a 4.7\% increase in the hazard. 

Note that those are practically significant effects, for two reasons. First, each coefficient provides the estimated relative increase in the hazard of becoming a team leader in some month $t$, given that the assumed behavior is exhibited at that same $t$. Conditional on the event not happening at $t$, the same behavior exhibited at $t+1$ will again lead to the same hazard increase. Second, because our dataset is dominated by zeros (i.e., our variables are distributed as power laws), a one standard deviation increase in any $X_{i}(t)$ actually corresponds to relatively small changes in behavior. For instance, as can be seen from Table~\ref{tab:explanatory_variables}, a one standard deviation increase in the number of pull requests submitted corresponds to 0.1 pull requests. In the case of focusfuture, it corresponds to a 0.036\% increase in the proportion of forward looking words.

\subsection{Leader emergence: simulations and counterfactuals} \label{survival}

\begin{figure}[ht]
     \centering
     \includegraphics[width=0.9\linewidth]{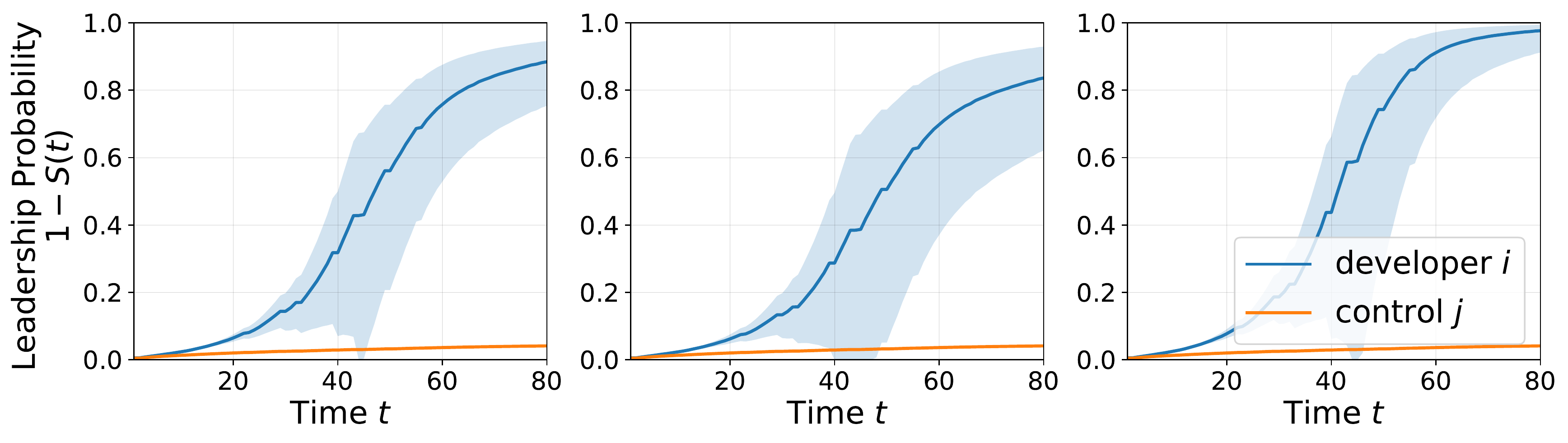}
     \caption{The orange line depicts the evolution of the leadership probability for a baseline developer $j$, inactive at all $t$. \textbf{Left}: developer $i$ is a ``fully inspirational'' developer, who makes no technical contributions. \textbf{Middle}: developer $i$ is a ``fully technical'' developer, who does not use inspirational communication. \textbf{Right}: developer $i$ combines the characteristics of the ``fully inspirational'' and ``fully technical'' developers. See the main text for details.}
     \label{fig:survial_curves_comms_pr}
\end{figure}

As explained in Section \ref{empirical_strategy}, one way to think about our estimated effect sizes is to consider developers' survival probability $S(t)$, i.e., the probability that the leadership event does not occur by time $t$ as a function of our covariates. More precisely, were are interested  in $1-S(t)$: the probability that the event \textit{does} occur given $\boldsymbol{X}(t)$. For illustration purposes, Figure~\ref{fig:survial_curves_comms_pr} reports the evolution of this probability over the time period covered by our study for three hypothetical developers (in blue), against a baseline developer whose covariates are fixed at zero over all $t$ (in orange). The developer on the left is assumed to make no technical contributions over all $t$, but lies in the 99.95 quantile over all $t$ in terms of the inspirational communication variables. The developer in the middle is assumed not to use inspirational communication over all $t$, but lies in the 99.95 quantile over all $t$ in terms of the technical contribution variables. Finally, the developer on the right is assumed to combine the characteristics of the left and middle developers, i.e., exhibit both strong technical and inspirational leadership.\footnote{Of course, those graphs should be understood as the result of a thought experiment concerning hypothetical developers. Because monthly participation is highly skewed at the developer level, the 99.95 quantile corresponds to sensible numbers for a highly active individual. Specifically, for the technical contribution variables, this quantile corresponds to 2 pull requests submitted per month, 7 issues raised, and 26 inline code comments received. (We set the technical leadership variables that have a negative impact on the hazard to 0.) For the inspirational communication variables, the quantile corresponds to \textit{prop(focusfuture)}=0.67\%, \textit{prop(we)}=0.38\%, \textit{prop(posemo)}=1.85\%, \textit{prop(negemo)}=0.58\%, and \textit{prop(certain)}=0.91\%. (We do not consider \textit{prop(tentat)}, since it has a statistically insignificant effect on the hazard.) For comparability, all models assume that the developer communicates at a rate of 400 words per month.} We can see that, according to our model, the leadership probabilities would evolve in similar ways for a ``fully inspirational'' and a ``fully technical'' developer, suggesting that both are equally important for leader emergence. In both cases, the leadership probability also converges to one by the end of our time period. Naturally, combining those characteristics in the right-hand side graph yields a steeper leadership probability curve.

In addition to simulating the process of leader emergence for theoretically interesting but hypothetical leader profiles, one can also use our model to predict the typical leader emergence process of a highly active leader or regular developer according to their observed behavior. To run this counterfactual exercise, we compare the predicted leader emergence process of a highly active leader or developer who respectively lies in the 99.95th quantile in terms of their technical and inspirational leadership variables over all $t$. In order to see whether highly active leaders within the most renowned projects exhibit different emergence trajectories, we further distinguish the behavior of leaders and developers within the top 100 Github projects (as measured by the number of stars received, a proxy for project success and popularity), from that of the rest of the Github population. 

\begin{figure}[ht]
     \centering
     \includegraphics[width=0.87\linewidth]{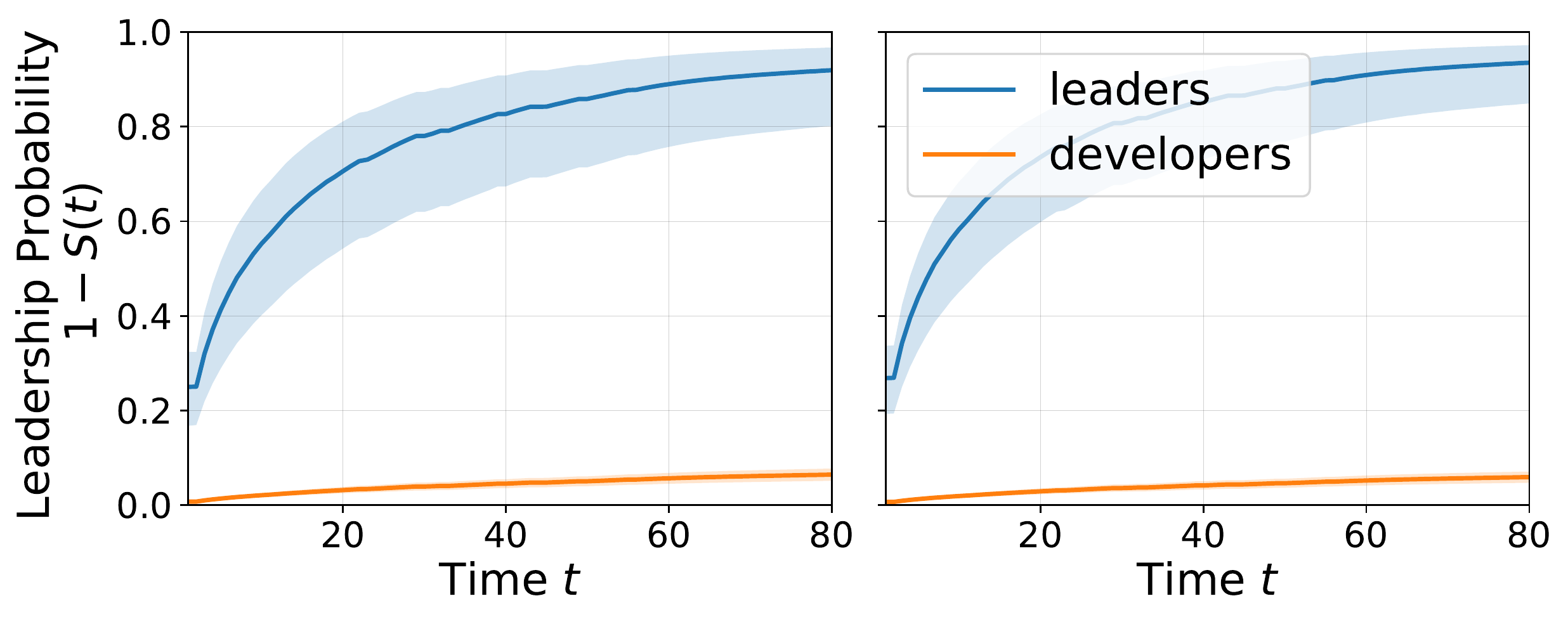}
     \caption{\textbf{Left}: simulated leader emergence probability for a highly active leader versus a highly active non-leader from one of the top 100 Github teams by reputation. \textbf{Right}: same as Left but for all remaining Github teams. See the main text for details.}
     \label{fig:surv_curves_top_teams}
\end{figure}

Figure~\ref{fig:surv_curves_top_teams} summarizes our results. The left graph predicts the average leader emergence process of a highly active leader (in blue) and developer (in orange) within a top 100 Github project, whereas the right graph does the same for all remaining projects. Two important conclusions follow from those leader emergence graphs. First, highly active leaders behave in a way that generates similar leader emergence trajectories, irrespective of whether they operate in a high status project or not. Second, developers who eventually become leaders behave in a distinct way, which translates into a predicted leadership probability that quickly converges to 1. By contrast, the predicted leader emergence process of highly active non-leaders is significantly slower in both graphs, and converges to about 10\% only by the end of our time period. 

\section{Conclusion}

We rely on the full record of developers' interactions with their teammates and projects at \texttt{github.com} between January 2010 and April 2017 to perform the first comprehensive study of the factors which predict leader emergence in open source software virtual teams. We organize our inquiry around two long standing narratives about the nature of leadership in open source, each related to independent streams of the theoretical literature on leadership in the social sciences -- one in economics, the other in management ~\citep{zehnder2017productive}. On the one hand, open source would work as a pure ``technical meritocracy'' ~\citep{raymond1999cathedral,weber2004success}. On the other, aspiring leaders would need to adopt ``inspirational'' communication strategies in order to gain and retain informal authority within their respective communities ~\citep{fogel2005producing,bacon2012art}. We set-up our study so as a to test for the empirical validity of the technical and inspirational leadership views when it comes to leader emergence in an online voluntary working space such as OSS. 

We find that, above and beyond developers' technical contributions and achievements, communication and community building skills emerge as strong predictors of leader emergence. The leadership preferences revealed by the open source community therefore support both the technical and inspirational leadership views. This result should be of interest to researchers and practitioners theorizing about OSS in particular and, more generally, leadership in geographically dispersed virtual team. They should also be of practical interest to online community managers seeking to earn and maintain leadership over peer production networks. 

Of course, OSS only represents one specific team production setting among many others, and future research should seek to compare our results to those obtained from other virtual or physical work settings. However, we do believe that open source provides a clean and highly ecologically valid opportunity to study the behavioral characteristics of \textit{naturally emerging} team leaders, for at least two reasons. First and foremost, leaders within open source software communities can only establish and maintain their authority through peer recognition. This is unlike many other team production environments (e.g., in firms or public organizations), where the skills necessary to arrive at leading positions can include institutional and political abilities as well as interpersonal connections, all of which are difficult to observe and control for. Second, OSS readily provides comprehensive public records of individuals' interactions with their respective teammates and projects that researchers can leverage to conduct quantitative analyses. Such activity traces are often missing from other team production settings, either because organizations are reluctant to share them, or because such records simply do not exist. 

We hope that the present paper will encourage further quantitative research on leadership in the context of geographically dispersed virtual teams. For instance, one important follow-on question that remains unaddressed by the literature relates to the effect that inspirational leaders may have on the effectiveness of their teams. In this paper, we sought to identify the behavioral correlates of endogenous leadership emergence in OSS and reveal the preferences of the extended community of developers in terms of leader characteristics. But if, all else equal, individuals prefer to appoint inspirational leaders, does that imply that inspirational leadership is causally linked to team performance? If so, not only should online community managers take notice of the preferences we reveal, but platform designers should also invest more heavily in the development and testing of features that may facilitate inspirational communication strategies and, as a result, promote cooperation and success at the team level. We leave this question (and others) open for future research.

\bigskip

\begin{footnotesize}

\noindent \textbf{Acknowledgements}

\noindent We are grateful to the Cooperation group at the Berkman Klein Center for Internet \& Society at Harvard University for insightful comments at the early stage of this project. We thank the ETH Zurich Scientific IT services, and, in particular, Emanuel Schmid, for indefectible logistical support. We thank Sai Aitharaju and Shiyi Li for top notch research assistance. We gratefully acknowledge financial support from the Center for Law and Economics at ETH Zurich, the ETH Zurich Career Seed grant, and the University of Strasbourg Attractivity grant.

\end{footnotesize}

\bigskip


\clearpage

\appendix{\Large{\textbf{APPENDIX}}}
\setcounter{figure}{0} \renewcommand{\thefigure}{A\arabic{figure}}
\setcounter{table}{0} \renewcommand{\thetable}{A\arabic{table}}
\smallskip

\section{Pairwise correlation between variables} \label{pwcorr}

\begin{figure}[h]
 \centering
 \hspace*{-1.1cm}
 \includegraphics[width=1\linewidth]{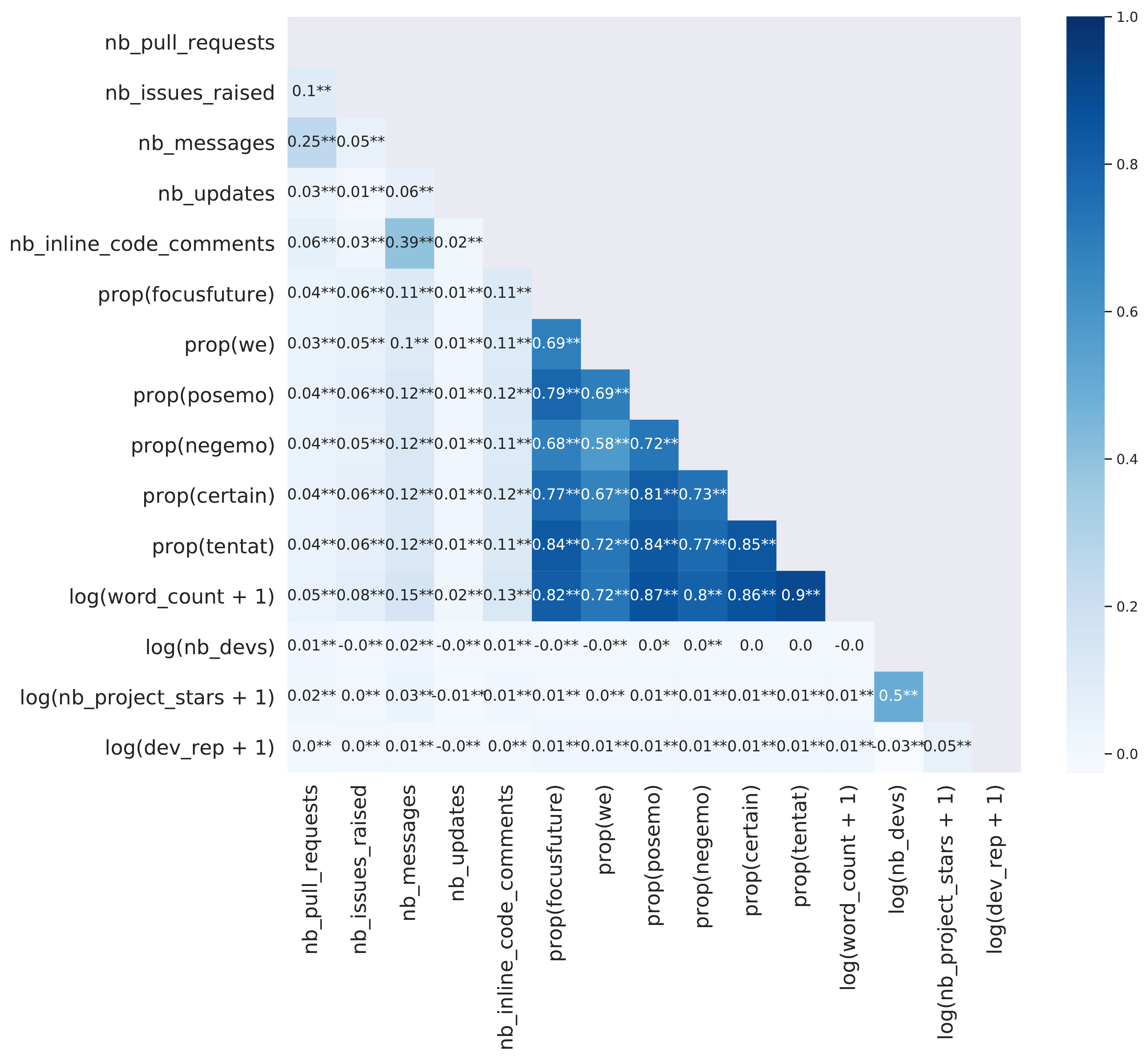}
 \caption{Pairwise correlation matrix of the variables used in the Extended Cox regressions. * and ** denote statistical significance with $p<0.01$ and $p<0.001$, respectively.}
 \label{fig:correlation_matrix}
\end{figure}

Figure ~\ref{fig:correlation_matrix} represents the pairwise correlation matrix between all the right hand side variables used in our model. The most notable feature of this table is the very strong pairwise correlation between all inspirational communication variables, which suggests that they move together at the developer level. The pairwise correlations between the technical leadership variables are relatively less strong, suggesting that each tends to capture a somewhat different dimension of technical expertise. Note that multicollinearity does not affect our point-estimates. Rather, it translates into higher confidence intervals, thus resulting in a relative loss of statistical power \citep{goldberger1991course,wooldridge2010econometric}. This inference problem can be solved by increasing the sample size, leading to more precise estimates. In the present case, the important size of our dataset allows us to estimate our effects precisely, notwithstanding the relatively high pairwise correlation between some of our variables.

\section{Leader emergence by quartile of team size} \label{q4results}

In this section, we reproduce the survival analysis reported in the paper, but check for potentially heterogeneous effects by segmenting the data in quartiles according to team size. As a result, this appendix presents four separate regressions focused on the following sub-groups of projects: 

\begin{enumerate}
    \item the first quartile contains projects with 3 developers or less,
    \item the second quartile contains projects with exactly 4 developers, 
    \item the third quartile contains projects with 5 to 10 developers, 
    \item the fourth quartile contains projects with more than 10 developers.
\end{enumerate}

Since our main analysis focuses on average effects within a population of relatively large projects (i.e., those that include a least 6 developers) this complementary analysis allows us to broadly confirm the generality of our results from Table \ref{table:result}. However, they also reveal some interesting heterogeneity in the leader emergence process according to team size. 

Most notably, it appears from a comparison of Tables \ref{table:q1}--\ref{table:q3} and Table \ref{table:q4} that the negative relationship uncovered in Table \ref{table:result} between leadership emergence prospects and (i) the monthly number of messages posted, and (ii) the number of follow-on updates to one's pull requests is in fact driven by the largest projects (i.e., those lying in the top quartile of team size and which include more than 10 developers). For all the smaller projects, both variables are positively associated with leader emergence. This result can be rationalized in terms of increasing coordination costs as the team gets large: posting many messages (for a given volume of text) as well as submitting code of presumably lower initial quality (i.e., code that requires several updates) may add net value to a small organization, but not to a larger one, where relatively inefficient communication and code contributions can become collectively costly on balance.

\section{Robustness of the sentiment analysis: NLTK VADER scores} \label{vader}

While the LIWC method for measuring sentiments in written text has been praised for its reliability and computational efficiency, even in Internet-mediated written exchanges ~\citep{golder2011diurnal,kivran2011network,de2013understanding,iosub2014emotions,pletea2014security}, its dictionaries have also sometimes been criticized for not sufficiently taking into account the specificity of social media communication, such as emoticons (e.g., ";)"), sentiment-ladden accronyms (e.g., "WTF"), slang (e.g., "the bomb"), and punctuation (e.g., "this is VERY good!!!!!"). 

In this section, we explore the robustness of our LIWC sentiment results by comparing them to those obtained using a different sentiment detection method: the NLTK Valence Aware Dictionary and sEntiment Reasoner (VADER). This method, itself hosted as a Github project\footnote{See \href{https://github.com/cjhutto/vaderSentiment}{https://github.com/cjhutto/vaderSentiment}}, has been developed precisely so as to detect a wider range of sentiments expressed through social media \citep{hutto2014vader}. 

\begin{figure}
     \centering
     \includegraphics[width=0.5\linewidth]{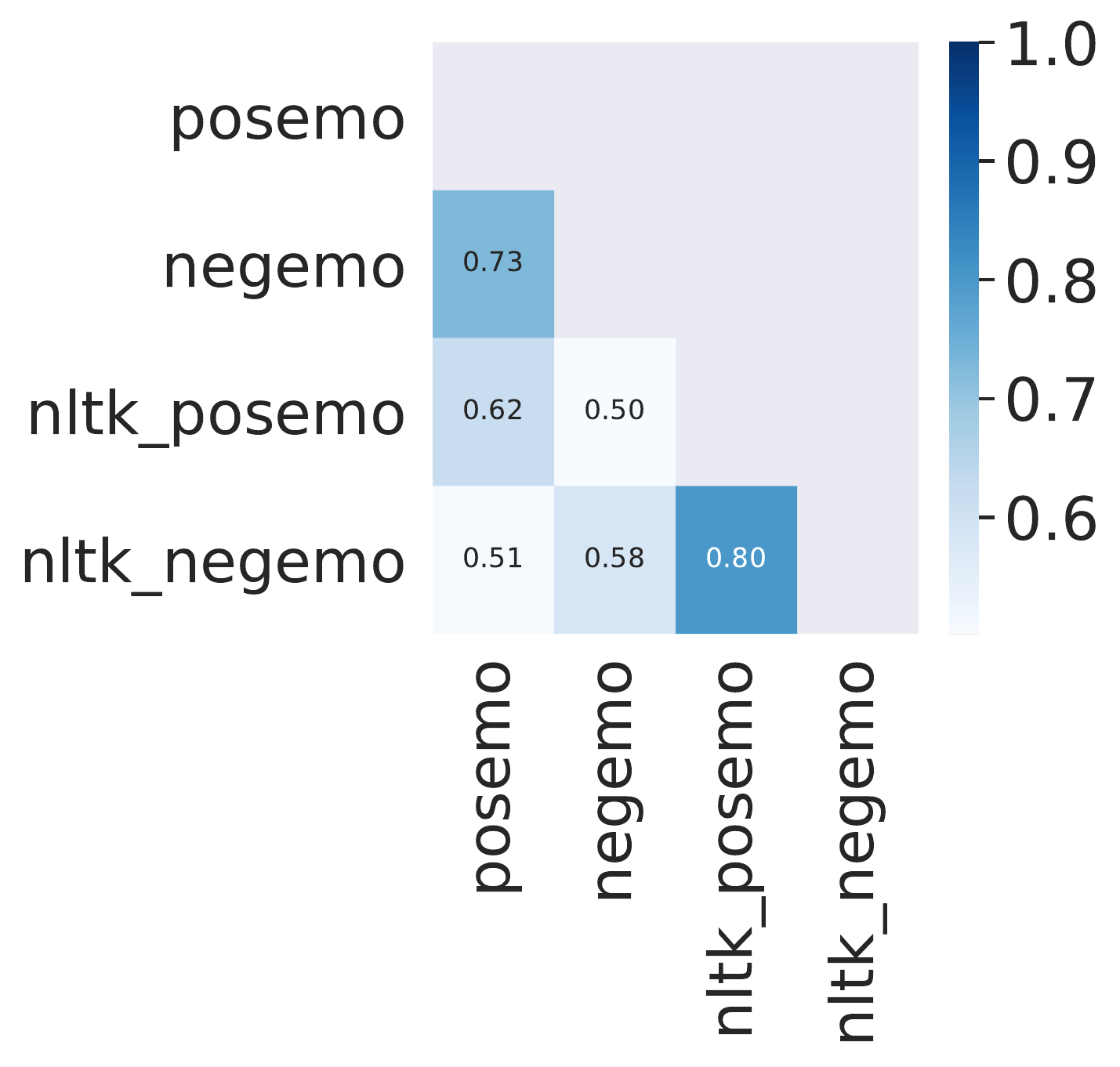}
     \caption{Correlation matrix between the LIWC positive and negative sentiment scores and the NLTK VADER scores.}
     \label{fig:nltk_liwc_q4_corr}
\end{figure}

We begin our analysis by reporting the pairwise correlation between the LIWC and NLTK VADER positive and negative sentiments scores in Figure~\ref{fig:nltk_liwc_q4_corr}. We can see that both methods deliver highly consistent results: the correlation between the LIWC and NLTK VADER sentiment scores is as high as 0.62 for positive emotions and 0.58 for negative emotions.

Next, we compare the sentiment coefficients obtained with both methods by replacing the LIWC positive emotion and negative emotion scores with their NLTK VADER equivalents. Since the NLTK VADER method is highly computationally intensive, we use a random subset of our data to perform this test. That is, we run our main regression with both LIWC and NLTK VADER sentiment scores using a quarter of the observations contained in the top quartile of projects in terms of team size. We report the results in Table~\ref{table:q4_minus} and \ref{table:q4_minus_nltk}, respectively. Consistent with the strong pairwise correlation between the LIWC and NLTK VADER sentiment scores, we can see that both methods deliver very similar results.

\begin{table*}
\caption{Correlates of leader emergence in OSS teams for the first quartile: $25\%-$ of team size. The table presents Extended Cox regression coefficients with robust standard errors clustered at the developer level. All variables are standardized apart from the control variables. This is done so that the magnitude of the coefficients can be directly compared and interpreted as the effect on the hazard of moving ``one standard deviation away'' from the sample mean for the variable considered at time $t$.}
\centering
\begin{tabular}{lccc} 
 \toprule
Variable & $\exp(\hat{\beta_j})$ & robust standard error & $p$-value \\
 \midrule
\textit{Technical leadership} \\
\text{ } nb pull requests (PR) & 1.006 & 0.001 & 2.32e-11 \\
\text{ } nb issues raised & 1.004 & 0.001 & < 2e-16 \\
\text{ } nb messages (PR \& issues) & 1.003 & 0.001 & 1.11e-08 \\
\text{ } nb of updates on PR & 1.006 & 0.001 & 9.24e-10 \\
\text{ } nb inline code comments & 1.010 & 0.001 &  < 2e-16 \\
\textit{Inspirational Leadership} \\
\text{ } prop(focusfuture) & 1.003 & 0.001 & 0.0318 \\
\text{ } prop(we) & 1.004 & 0.001 & 1.14e-05 \\
\text{ } prop(posemo) & 1.008 & 0.002 & 4.94e-06 \\
\text{ } prop(negemo) & 1.002 & 0.001 & 0.1701 \\
\text{ } prop(certain) & 1.002 & 0.002 & 0.1923 \\
\text{ } prop(tentative) & 0.999 & 0.002 & 0.4461 \\
\textit{Controls} \\
\text{ } log(1 + word count) & 2.071 & 0.018 & < 2e-16 \\
\text{ } log(nb of developers in team) & 1.564 & 0.054 & < 2e-16 \\
\text{ } log(1 + nb project stars) & 1.165 & 0.015 & < 2e-16 \\
\text{ } log(1 + developer reputation) & 1.060 & 0.011 & 1.32e-07 \\
\midrule
nb observations = 12,193,437  \\
nb events = 46,339 \\
\bottomrule
\end{tabular}
\label{table:q1}
\end{table*}

\begin{table*}
\caption{Correlates of leader emergence in OSS teams for the second quartile: $25\%$ to $50\%$ of team size. The table presents Extended Cox regression coefficients with robust standard errors clustered at the developer level. All variables are standardized apart from the control variables. This is done so that the magnitude of the coefficients can be directly compared and interpreted as the effect on the hazard of moving ``one standard deviation away'' from the sample mean for the variable considered at time $t$.}
\centering
\begin{tabular}{lccc} 
 \toprule
Variable & $\exp(\hat{\beta_j})$ & robust standard error & $p$-value \\
 \midrule
\textit{Technical leadership} \\
\text{ } nb pull requests (PR) & 1.006 & 0.002 & 0.01263 \\
\text{ } nb issues raised & 1.009 & 0.001 & 2.95e-14 \\
\text{ } nb messages (PR \& issues) & 1.005 & 0.001 & 8.34e-07 \\
\text{ } nb of updates on PR & 1.008 & 0.001 & 2.69e-10 \\
\text{ } nb inline code comments & 1.007 & 0.001 &  1.79e-10 \\
\textit{Inspirational Leadership} \\
\text{ } prop(focusfuture) & 1.001 & 0.002 & 0.58229 \\
\text{ } prop(we) & 1.007 & 0.002 & 0.00101 \\
\text{ } prop(posemo) & 1.011 & 0.003 & 0.00132 \\
\text{ } prop(negemo) & 1.000 & 0.002 & 0.84035 \\
\text{ } prop(certain) & 1.001 & 0.003 & 0.82894 \\
\text{ } prop(tentative) & 1.007 & 0.003 & 0.04207 \\
\textit{Controls} \\
\text{ } log(1 + word count) & 2.007 & 0.017 & < 2e-16 \\
\text{ } log(nb of developers in team) & NA & NA & NA \\
\text{ } log(1 + nb project stars) & 1.079 & 0.003 & 2.24e-08 \\
\text{ } log(1 + developer reputation) & 1.033 & 0.0075 & 1.81e-05 \\
\midrule
nb observations = 4,710,354  \\
nb events = 24,674 \\
\bottomrule
\end{tabular}
\label{table:q2}
\end{table*}

\begin{table*}
\caption{Correlates of leader emergence in OSS teams for the third quartile: $50\%$ to $75\%$ of team size. The table presents Extended Cox regression coefficients with robust standard errors clustered at the developer level. All variables are standardized apart from the control variables. This is done so that the magnitude of the coefficients can be directly compared and interpreted as the effect on the hazard of moving ``one standard deviation away'' from the sample mean for the variable considered at time $t$.}
\centering
\begin{tabular}{lccc} 
 \toprule
Variable & $\exp(\hat{\beta_j})$ & robust standard error & $p$-value \\
 \midrule
\textit{Technical leadership} \\
\text{ } nb pull requests (PR) & 1.006 & 0.003 & 0.020822 \\
\text{ } nb issues raised & 1.011 & 0.001 & < 2e-16 \\
\text{ } nb messages (PR \& issues) & 1.006 & 0.002 & 0.000126 \\
\text{ } nb of updates on PR & 1.008 & 0.001 & 1.43e-09 \\
\text{ } nb inline code comments & 1.004 & 0.001 &  4.57e-12 \\
\textit{Inspirational Leadership} \\
\text{ } prop(focusfuture) & 1.006 & 0.002 & 0.000153 \\
\text{ } prop(we) & 1.010 & 0.0012 & 1.01e-15 \\
\text{ } prop(posemo) & 1.015 & 0.002 & 2.18e-14 \\
\text{ } prop(negemo) & 1.004 & 0.001 & 0.008444 \\
\text{ } prop(certain) & 1.008 & 0.002 & 2.20e-05 \\
\text{ } prop(tentative) & 1.002 & 0.002 & 0.435437 \\
\textit{Controls} \\
\text{ } log(1 + word count) & 1.855 & 0.009 & < 2e-16 \\
\text{ } log(nb of developers in team) & 1.057 & 0.025 & 0.025328 \\
\text{ } log(1 + nb project stars) & 1.033 & 0.008 & 8.35e-05 \\
\text{ } log(1 + developer reputation) & 1.014 & 0.005 & 0.012640 \\
\midrule
nb observations = 12,148,021  \\
nb events = 62,878 \\
\bottomrule
\end{tabular}
\label{table:q3}
\end{table*}

\begin{table*}
\caption{Correlates of leader emergence in OSS teams for the fourth quartile: $75\%+$ of team size. The table presents Extended Cox regression coefficients with robust standard errors clustered at the developer level. All variables are standardized apart from the control variables. This done so that the magnitude of the coefficients can be directly compared and interpreted as the effect on the hazard of moving ``one standard deviation away'' from the sample mean for the variable considered at time $t$.}
\centering
\begin{tabular}{lccc} 
 \toprule
Variable & $\exp(\hat{\beta_j})$ & robust standard error & $p$-value \\
 \midrule
\textit{Technical leadership} \\
\text{ } nb pull requests (PR) & 1.016  &  0.001 & < 2e-16\\
\text{ } nb issues raised & 1.005  &  0.000 & < 2e-16 \\
\text{ } nb messages (PR \& issues) & 0.977  &  0.002 & < 2e-16 \\
\text{ } nb of updates on PR & 0.994  &  0.003 & 0.061859 \\
\text{ } nb inline code comments & 1.014  &  0.002 &  1.15e-12 \\
\textit{Inspirational Leadership} \\
\text{ } prop(focusfuture) & 1.013  &  0.001 & < 2e-16 \\
\text{ } prop(we) & 1.013  &  0.001 & < 2e-16 \\
\text{ } prop(posemo) & 1.003  &  0.001 & 0.032667 \\
\text{ } prop(negemo) & 1.007  &  0.001 & 9.87e-08 \\
\text{ } prop(certain) & 1.012  &  0.001 & < 2e-16 \\
\text{ } prop(tentative) & 1.006  &  0.002 & 0.000792 \\
\textit{Controls} \\
\text{ } log(1 + word count) & 1.877  &  0.006 & < 2e-16 \\
\text{ } log(nb of developers in team) & 0.748  &  0.007 & < 2e-16 \\
\text{ } log(1 + nb project stars) & 1.119  &  0.003 & < 2e-16 \\
\text{ } log(1 + developer reputation) & 1.006  &  0.003 & 0.044167 \\
\midrule
nb observations = 22,926,156  \\
nb events = 75,625 \\
\bottomrule
\end{tabular}
\label{table:q4}
\end{table*}

\begin{table*}
\caption{Correlates of leader emergence in OSS teams using LIWC sentiment scores. The regression is based on a random sample of data extracted from the population of projects lying in the top quartile in terms of team size. The table presents Extended Cox regression coefficients with robust standard errors clustered at the developer level. All variables are standardized apart from the control variables. This done so that the magnitude of the coefficients can be directly compared and interpreted as the effect on the hazard of moving ``one standard deviation away'' from the sample mean for the variable considered at time $t$.}
\centering
\begin{tabular}{lccc} 
 \toprule
Variable & $\exp(\hat{\beta_j})$ & robust standard error & $p$-value \\
 \midrule
\textit{Technical leadership} \\
\text{ } nb pull requests (PR) & 1.028  & 0.002 & < 2e-16 \\
\text{ } nb issues raised & 1.005  & 0.001 & < 2e-16 \\
\text{ } nb messages (PR \& issues) & 1.015 &  0.004 & 0.000113 \\
\text{ } nb of updates on PR & 1.003  & 0.003 & 0.399  \\
\text{ } nb inline code comments & 1.007  & 0.003 & 0.00933 \\
\textit{Inspirational Leadership} \\
\text{ } prop(focusfuture) & 1.008 & 0.002 & 0.000775 \\
\text{ } prop(we) & 1.017 &  0.002 & < 2e-16 \\
\text{ } prop(posemo) & 1.016  &  0.003 & 1.63e-08 \\
\text{ } prop(negemo) & 1.004 &  0.002 & 0.106 \\
\text{ } prop(certain) & 1.007 & 0.003 & 0.0116\\
\text{ } prop(tentative) & 1.002  &  0.004 & 0.4852 \\
\textit{Controls} \\
\text{ } log(1 + word count) & 1.650 & 0.007 & < 2e-16 \\
\text{ } log(nb of developers in team) & 0.811 & 0.005 & < 2e-16 \\
\text{ } log(1 + nb project stars) & 1.052 &  0.003 & < 2e-16 \\
\text{ } log(1 + developer reputation) & 1.001 & 0.002 & 0.762248 \\
\midrule
nb observations = 5,636,515 \\
nb events = 42,191 \\
\bottomrule
\end{tabular}
\label{table:q4_minus}
\end{table*}

\begin{table*}
\caption{Correlates of leader emergence in OSS teams using NLTK VADER sentiment scores. The regression is based on a random sample of data extracted from the population of projects lying in the top quartile in terms of team size. The table presents Extended Cox regression coefficients with robust standard errors clustered at the developer level. All variables are standardized apart from the control variables. This done so that the magnitude of the coefficients can be directly compared and interpreted as the effect on the hazard of moving ``one standard deviation away'' from the sample mean for the variable considered at time $t$.}
\centering
\begin{tabular}{lccc} 
 \toprule
Variable & $\exp(\hat{\beta_j})$ & robust standard error & $p$-value \\
 \midrule
\textit{Technical leadership} \\
\text{ } nb pull requests (PR) & 1.028  & 0.002 & < 2e-16 \\
\text{ } nb issues raised & 1.005  & 0.000 & < 2e-16 \\
\text{ } nb messages (PR \& issues) & 1.014  & 0.004 & 0.000230 \\
\text{ } nb of updates on PR & 1.002 & 0.003 &  0.474810  \\
\text{ } nb inline code comments & 1.007 & 0.003 & 0.009509 \\
\textit{Inspirational Leadership} \\
\text{ } prop(focusfuture) & 1.009 & 0.002 & 0.000193 \\
\text{ } prop(we) & 1.017 &  0.002 & < 2e-16 \\
\text{ } nltk prop(posemo) & 1.008  & 0.002 & 0.001547 \\
\text{ } nltk prop(negemo) & 1.004 &  0.002 & 0.068100 \\
\text{ } prop(certain) & 1.010  &  0.003 & 0.000410 \\
\text{ } prop(tentative) & 1.006  & 0.003 & 0.092221 \\
\textit{Controls} \\
\text{ } log(1 + word count) & 1.654 & 0.007 & < 2e-16 \\
\text{ } log(nb of developers in team) & 0.812 & 0.005 & < 2e-16 \\
\text{ } log(1 + nb project stars) & 1.053 &  0.003 & < 2e-16 \\
\text{ } log(1 + developer reputation) & 1.000 & 0.002 & 0.906157\\
\midrule
nb observations = 5,636,515\\
nb events = 42,191 \\
\bottomrule
\end{tabular}
\label{table:q4_minus_nltk}
\end{table*}

\end{document}